\documentclass[11pt,tightenlines,eqsecnum,floats,aps,amssymb,nofootinbib,prd,shownopacs,floatfix,superscriptaddress]{revtex4-2}

\usepackage{graphicx}
\usepackage{epstopdf}
\usepackage{latexsym}
\usepackage{amssymb}
\usepackage{amsmath}
\usepackage{color}
\usepackage{mathrsfs}
\usepackage{xparse}
\usepackage[dvipsnames]{xcolor}
\usepackage{float}
\usepackage[colorinlistoftodos,prependcaption,textsize=tiny]{todonotes}
\usepackage{dsfont}
\usepackage{mathtools}
\usepackage[toc,title,page]{appendix}
\usepackage[colorlinks=false, pdfborder={0 0 0}]{hyperref}
\usepackage{physics}

\usepackage{tensor}
\usepackage{braket}

\numberwithin{equation}{section}

\usepackage[colorinlistoftodos,prependcaption,textsize=tiny]{todonotes}

\begin{document}

\title{Formation of shell-crossing singularities in effective gravitational collapse models with bounded and unbounded polymerizations}

\author{Francesco Fazzini} \email{francesco.fazzini@fau.de}

\author{Kristina Giesel} \email{kristina.giesel@fau.de}

\author{Eric Rullit} \email{eric.rullit@fau.de}

\affiliation{Institute for Quantum Gravity, Theoretical Physics III, Department of Physics,  Friedrich-Alexander-Universit\"at Erlangen-N\"urnberg, Staudtstr. 7, 91058 Erlangen, Germany.}

\begin{abstract}
\noindent
We extend the investigation into the formation of shell-crossing singularites (SCS) in effective polymerized LTB models to the LQG-inspired asymmetric bounce model, as well as to effective LTB models based on the solutions of Bardeen and Hayward, in which no bounce occurs. While the asymmetric bouncing model belongs to the class of bounded polymerization functions, the latter models feature unbounded polymerization functions.
 Our results show that, similar to the symmetric bouncing model, for the asymmetric bouncing model SCS are unavoidable for inhomogeneous dust profiles. In contrast, for models without a bounce and with unbounded polymerization functions, no SCS form for inhomogeneous, decreasing dust profiles—a situation that resembles classical theory, in which SCS can also be avoided by a suitable choice of initial data.
\end{abstract}

\maketitle

\section{Introduction}
\label{sec:Intro}
\noindent Black holes represent an important prediction for the verification of classical General Relativity (GR). Long supported by four decades of indirect observations \cite{LIGOScientific:2016aoc,1972Natur.235...37W,Ghez:2008ms}, their existence has been confirmed through direct imaging \cite{EventHorizonTelescope:2019dse} by the Event Horizon Telescope collaboration, and their merger detection by LIGO and VIRGO collaborations \cite{LIGOScientific:2016aoc}, providing a robust experimental foundation for the theory. The primary mechanism for black hole formation is well-established: the gravitational collapse of a massive star under its own weight, conceptually pionered by the work of Oppenheimer and Snyder \cite{Oppenheimer:1939ue}.
At the theoretical level, a simple yet profound realization of this process is provided by the Lemaître-Tolman-Bondi (LTB) spacetime \cite{Lemaitre:1933gd,Tolman:1934za,
Bondi:1947fta}---of which the Oppenheimer-Snyder model is a special case, representing the collapse of a homogeneous dust ball. Describing spherically symmetric (generally inhomogeneous) dust collapse, the LTB model tracks the evolution of an arbitrary initial density profile toward the formation of a black hole, ultimately leading to a central singularity. 
~\\
~\\
The presence of a central singularity is a common feature of many spacetimes in GR \cite{Penrose:1964wq}. At the central singularity, GR fails as a theory and loses its predictive power, since it offers no way to extend the spacetime manifold beyond such a point—a phenomenon also known as geodesic incompleteness. Furthermore, quantities such as the curvature scalars diverge at a central singularity, and ordinary matter is compressed to infinite density by tidal forces—an effect often referred to as “spaghettification”—with these forces diverging as the singularity is approached \cite{Joshi:2024djy}.

Despite its prominence, the central singularity is not the only divergence that can emerge in LTB spacetimes. Another source of physical divergence for curvature scalars due to matter density divergence is the formation of shell-crossing singularities (SCS) \cite{Szekeres:1995gy}. These occur when distinct matter layers, traveling at different relative velocities, cross one another, leading to a divergence in the volumetric energy density and the formation of an infinitesimally thin shell of dust. Although there is currently no known classical mechanism to systematically avoid these singularities (even the inclusion of pressure does not guarantee their systematic avoidance \cite{hagendas1974}), they are considered less pathological than the central singularity for several reasons. First, while curvature scalars and radial tidal forces diverge at a SCS, angular tidal forces remain finite. More importantly, the radial deviation of geodesics remains finite, leading to a less dramatic spaghettification effect \cite{Szekeres:1995gy}. Second, as demonstrated in \cite{Hellaby:1985zz}, within the classical context, there exists a broad class of initial conditions that do not develop SCS prior to the central singularity. Finally, due to their weaker nature, extensions of the spacetime beyond SCS are possible through the use of weak solutions \cite{Nolan:2003wp,Husain:2025wrh} or the Israel junction conditions \cite{Maeda1983} (studied in a cosmological inhomogeneous scenario). For a comparison between the two approaches, see \cite{Fazzini:2025zrq}.
~\\
~\\
With the classical picture in mind, the primary objective of this work is to examine how this scenario is altered when quantum gravitational effects are taken into account. Indeed, the resolution of classical singularities is a central motivation for the development of a quantum theory of gravity. Although a complete model of gravitational collapse derived directly from a fundamental quantum theory is currently unavailable, there exists a diverse range of effective models (see e.g. \cite{Lewandowski:2022zce,Husain:2022gwp,Bobula:2023kbo,Alonso-Bardaji:2023qgu,Bojowald:2024ium,Han:2023wxg,Cafaro:2024vrw,Liu:2025fil}).
In such models, the quantum gravitational corrections are encoded in so-called polymerization functions, which modify the form of the classical dynamics and are expected to provide a suitable effective description of the underlying quantum dynamics of the symmetry reduced models \cite{Ashtekar:2005qt,Modesto:2005zm,Boehmer:2007ket,Chiou:2012pg,Gambini:2013hna,Brahma:2014gca,Dadhich:2015ora,Tibrewala:2013kba,BenAchour:2017ivq,Yonika:2017qgo,DAmbrosio:2020mut,Olmedo:2017lvt,Ashtekar:2018lag,Ashtekar:2018cay,Bojowald:2018xxu,BenAchour:2018khr,Bodendorfer:2019cyv,Alesci:2019pbs,Assanioussi:2019twp,Benitez:2020szx,Gan:2020dkb,Gambini:2020qhx,Husain:2021ojz,Husain:2022gwp,Li:2021snn,Gan:2022mle,Kelly:2020uwj,Gambini:2020nsf,Han:2020uhb,Zhang:2021xoa,Munch:2022teq,Lewandowski:2022zce,Giesel:2021dug,Giesel:2022rxi,Han:2022rsx,Giesel:2023tsj,Giesel:2024mps,Cafaro:2024vrw}. For cosmological models inspired by Loop Quantum Gravity (LQG), these effective models were derived either directly in the symmetry reduced sector \cite{Taveras:2008ke} or from the full LQG within the framework of Algebraic Quantum Gravity \cite {Giesel:2007wn} in \cite{Dapor:2017rwv}. A complete derivation in the context of spherically symmetric models is not available in the literature yet but rather currently being investigated in ongoing research. For the models considered in this article, the latter is unproblematic, since we are considering either LTB models constructed from an infinite number of decoupled cosmological models—one for each dust shell, following \cite{Husain:2025wrh,Lewandowski:2022zce,Giesel:2023tsj,Giesel:2026pjj}—or the models of Bardeen \cite{2767662} and Hayward \cite{Hayward:2005gi} that can be linked to (generlized) extended mimetic gravity models as the underlying theoretical framework, as discussed in \cite{Giesel:2024mps,Giesel:2026pjj}. These models are qualitatively different: the effective models inspired by LQG exhibit a bounce and are based on bounded polymerization functions, whereas the models by Bardeen \cite{2767662} and Hayward \cite{Hayward:2005gi} do not exhibit a bounce and involve unbounded polymerization functions. Despite their qualitative differences, all models are characterized by the resolution of the central singularity through quantum gravitational corrections that can be associated with certain extended mimetic gravitational models \cite{Giesel:2024mps}.
~\\
~\\
The primary focus of this work is the investigation of the  formation of SCS within these effective models, following the precursor analysis performed for a LQG inspired specific model with a symmetric bounce \cite{Fazzini:2023ova}. We will investigate three particular models in this work: for LQG inspired models we consider the one from \cite{Dapor:2017rwv} with an asymmetric bounce as an example for bounded polymerizations and the models of  Bardeen \cite{2767662} and Hayward as examples for unbounded polymerizations. This study has two objectives: first, it aims to investigate the role that current quantum gravitational corrections play in the formation of SCS and whether they can prevent such singularities. Second, it seeks to determine whether the conditions governing the presence or absence of SCS allow for a possible classification of these models within a broader perspective. 
~\\
The paper is structured as follows: after the introduction in section \ref{sec:Intro}, we briefly review the main properties of the effective LTB models with dust considered in this work. In section \ref{sec:FormSCS} we investigate the formation of SCS for the three different models introduced in section \ref{sec:ReviewEffModels}. We summarize our results and give an outlook for future steps in section \ref{sec:Conclusion}.   
\section{Effective LTB models with dust}\label{sec:ReviewEffModels}

In this section we provide a concise overview of the LTB models introduced in section \ref{sec:Intro} within the framework of effective LQG-inspired models, followed by the introdution of three concrete examples that will serve as the main focus of the subsequent analysis. To fix the notation, we work with Ashtekar-Barbero variables which adapted to the spherical symmetry reduction read $(K_x, E^x)$, $(K_\phi, E^\phi)$, satisfying standard Poisson brackets \cite{Husain:2022gwp, Giesel:2023hys}. Here, $K$ denotes the extrinsic curvature and $E$ the densitized triad. In addition, we consider canonical dust variables $(T, P_T)$ from the non-rotational dust model and use  $T$ as a dynamical reference field to define a phyiscal time evolution with respect to the comoving dust time, corresponding to the gauge condition $T- t = 0$. The general spherically symmetric line element written in terms of the above variables then has the form
\begin{align}\label{eq:LE1}
    \dd s^2 = -\dd t^2 + \frac{(E^\phi)^2}{|E^x|}(\dd x + N^x \dd t)^2 + |E^x|\dd\Omega^2,
\end{align}
where $(t, x, \theta, \phi)$ are comoving coordinates and $N^x$ is the radial component of the shift vector. In the same set of coordinates, LTB models are characterized by the line element \cite{Lemaitre:1933gd, Tolman:1934za, Bondi:1947fta}
\begin{align}\label{eq:LE2}
    \dd s^2 = -\dd t^2 + \frac{((E^x)^\prime)^2}{4|E^x|(1 + \mathcal{E}(x))} \dd x^2 + |E^x| \dd\Omega^2,
\end{align}
where $\mathcal{E}(x)$ denotes the LTB functions that can be understood as the total energy of the dust shell.  In particular, it is rediscovered from \eqref{eq:LE1} via the so-called LTB condition $|E^x|^\prime - 2 E^\phi \sqrt{1 + \mathcal{E}(x)} = 0$. We introduce the areal radius of the dust shell $R = R(t, x)=\sqrt{E^x}$  and using Einstein's field equations it follows that  $R$ satisfies the following evolution equation
\begin{align}\label{eq:FE}
    \Dot{R}^2 = \mathcal{E}(x) + \frac{2 G M(x)}{R},
\end{align}
where $G$ is the gravitational constant and $M(x)$ the Misner-Sharp mass measuring the total gravitational mass confined inside a shell. From the form of the evolution equations that for each shell has the form of a FLRW equation for $R$ we realize that the LTB function $\mathcal{E}(x)$ plays the role of a curvature profile.  Since no radial derivatives are involved in \eqref{eq:FE}, it can be interpreted as a set of infinitely many decoupled Friedmann equations for each individual dust shell labeled by the radial coordinate $x$.
~\\
~\\
Similar considerations apply to the class of effective models that will be analyzed in the following. These effective models are in general characterized by quantum gravitational corrections which are inspired from a LQG inspired quantization of the symmetry reduced sector. In particular, connection variables are replaced by corresponding polymerization functions and inverse triads are replaced by suitable corrections, see for instance \cite{Bojowald:2008ja, Bojowald:2009ih}. An approach followed in the works \cite{Giesel:2023tsj, Giesel:2024mps} and applied to a specific model in \cite{Giesel:2023hys} is to start with a general expression for the gravitational contribution of the effective Hamiltonian constraint in spherical symmetry by keeping the functions describing the two types of quantum corrections denoted as $f(K_x, K_\phi, E^x, E^\phi)$ and $h_1(E^x)$, $h_2(E^\phi)$ unspecified, which allows to cover a wide range of existing effective models. More general polymerization as for instance used in \cite{Alonso-Bardaji:2021yls} do not fall in the class of models considered in \cite{Giesel:2023tsj, Giesel:2024mps}. It was then shown that the requirement of additional conditions for the underlying model yields to restrictions on the polymerization functions. For instance, requiring that a compatible LTB reduction exist and that it is preserved under the effective dynamics rules out specific polymerizations. Similar yields the requirement that the spatial diffeomporphism constraint, which keeps its classical form in the models considered, and the effective Hamiltonian constraint form a closed Poisson algebra to further restrictions for the allowed polymerization functions. For our purposes we are primarily interested in models for which $i)$ a compatible LTB condition exists, $ii)$ the gravitational contribution of the effective Hamiltonian constraint is a conserved quantity for each shell and hence the effective dynamics completely decouples along the radial coordinate and $iii)$ the inverse triad corrections are absent due to the existence of an underlying mimetic Lagrangian for these models\footnote{The effective models meeting these conditions belong to the class II $\cap$ III in \cite{Giesel:2024mps}.}. The second condition implies in particular that the mass function $M(x)$ for each individual dust shell is a conserved quantity just as in the classical case, which provides a crucial simplification for studying the effective dynamics of dust collapsing models. In such a case the gravitational contribution of the effective Hamiltonian constraint restricted to the LTB sector can be written as \cite{Giesel:2023tsj, Giesel:2024mps}
\begin{align}\label{eq:ShellH}
    C^{(\alpha)}(x)_{\mathrm{LTB}} = \frac{\partial_x \widetilde{H}^{(\alpha)}(x)}{2G \sqrt{1 + \mathcal{E}(x)}},\hspace{.5cm}\widetilde{H}^{(\alpha)}(x) = M(x),
\end{align}
where $\widetilde{H}^{(\alpha)}(x)$ is a model specific shell Hamiltonian which contains the polymerization functions obeying the above requirements. The so-called polymerization parameter $\alpha$ encodes the magnitude of quantum effects and is subject to the classical limit $\alpha \to 0$ for which the effective shell dynamics following from \eqref{eq:ShellH} is required to return to its classical counterpart given by \eqref{eq:FE}. Accordingly, the evolution equation generated by $\widetilde{H}^{(\alpha)}(x)$ will be referred as the modified Friedmann equation.
~\\
~\\
As outlined in \cite{Giesel:2024mps}, the approach reviewed in the previous paragraph goes beyond the standard LQG-inspired modifications which are typically given in terms of bounded functions and controlled by a polymerization parameter $\alpha_\Delta = \gamma \sqrt{\Delta}$ associated to the minimal area gap $\Delta$ \cite{Rovelli:1994ge, Ashtekar:1996eg}. The formalism also accommodates a different class of effective models derived from given regular black hole solutions in vacuum such as Bardeen and Hayward which instead turn out to be characterized by unbounded monotonic polymerization functions. The asymmetric bounce model that has been investigated recently in \cite{Giesel:2026pjj} in the context of effective LTB models belongs the former case as it borrows its modifications from the Thiemann regularized LQC model derived in \cite{Yang:2009fp, Dapor:2017rwv}. That is, similar to the symmetric bouncing model extensively studied in the literature \cite{Giesel:2023hys,Husain:2022gwp,Fazzini:2023ova,Bobula:2024chr}, each individual shell is considered as a cosmological model by property $ii)$ and its dynamics is governed by the modified Friedmann equation \cite{Giesel:2024mps, Giesel:2026pjj}
\begin{align}
\label{eq:ModFRWAsymB}
    \Dot{R} = \pm \frac{x_0}{\sqrt{2} \alpha_\Delta ({\gamma}^2+1) R^2} \sqrt{\frac{\mp x_0 + 4 \alpha_\Delta^2 \left({\gamma}^2+1\right) G M(x)+R^3}{R^3}},
\end{align}
where $x_0 := [R^6-8 \alpha_\Delta^2 {\gamma}^2 \left({\gamma}^2+1\right) G R^3 M(x)]^{1/2}$. While the analytical solution on which the proceeding analysis will be build on is provided in section \ref{sec:A}, we can infer already at this stage that the solution admits a bouncing behaviour with a well-defined minimal areal radius following from the global sign change at $\Dot{R} = 0$.

Different from bouncing models with LQG-inspired modifications are the previously mentioned regular black hole solutions in spherical symmetry of which Bardeen \cite{2767662} and Hayward \cite{Hayward:2005gi} are well-known examples. These solutions are given in terms of line elements in Schwarzschild-like coordinates of the form
\begin{align}\label{eq:metric}
    \dd s^2 = -(1 - \mathcal{G}(r)^2) \dd \tau^2 + (1 - \mathcal{G}(r)^2)^{-1} \dd r^2 + r^2 \dd\Omega^2,
\end{align}
where $\mathcal{G}(r)$ denotes the metric function that contains modifications compared to the standard Schwarzschild metric. The effective LTB models are then constructed gradually following the so-called reconstruction algorithm introduced in \cite{Giesel:2024mps} which takes \eqref{eq:metric} as a starting point and in which the modifications are are linked to certain choice of polymerization functions. These in turn are used to obtain shell Hamiltonians that encode the dynamics of the resulting effective model by means of decoupled modified Friedmann equations for the areal radius $R$ of the individual dust shells. In fact, the latter can be directly obtained from the metric function once $\mathcal{G}(r)^2$ is expressed as a function $r_s / r^3$, that is $\mathcal{G}(r)^2 =r^2\widetilde{\mathcal{G}}(\frac{r_s}{r^3})^2$ with $r_s := 2Gm$. This amounts to redefining the modifications contained in $\mathcal{G}(r)^2$ in an suitable manner using the polymerisation parameter $\alpha$. Then the modified Friedmann equation can be obtained from the formula \cite{Giesel:2024mps, Giesel:2025kdl}
\begin{align}
    \Dot{R}^2 = \mathcal{E}(x) + R^2 \widetilde{\mathcal{G}}^2\left(\frac{2 G M(x)}{R^3}\right),
\end{align}
where we identified $r = R(t, x)$ and $m = M(x)$. As an example, the metric functions of the Bardeen and Hayward model are given by
\begin{align}
    \mathcal{G}_\mathrm{B}(r)^2 = \frac{r_s r^2}{(r^2 + l_1^2)^{3/2}},\hspace{.5cm}\mathcal{G}_\mathrm{H}(r)^2 = \frac{r_s r^2}{r^3 + l_2^2 r_s},
\end{align}
where $l_1$, $l_2$ denote the modifications of the standard metric function. After rewriting the modifications as $l_1 = \alpha^{2/3} r_s^{1/3}$ and $l_2 = \alpha$ in order for $\widetilde{\mathcal{G}}^2$ to be a function of $r_s / r^3$ in both cases, the modified Friedmann equations are computed to
\begin{align}
    \text{Bardeen:}\hspace{.3cm}\frac{\Dot{R}^2}{R^2} &= \frac{2 G M(x)}{\left[R^2 + (2 G M(x) \alpha^2)^{2/3}\right]^{3/2}}, \\
    \text{Hayward:}\hspace{.3cm}\frac{\Dot{R}^2}{R^2} &= \frac{2 G M(x)}{R^3 + 2 G M(x) \alpha^2},
\end{align}
with analytical solutions given in \eqref{eq:RBardeen} and \eqref{eq:RHayward} respectively. Note that we restricted to the marginally bound sector $\mathcal{E}(x) = 0$ for all three models. A discussion of the more general non-marginally bound case in which $\mathcal{E}(x) \neq 0$ can be found in \cite{Giesel:2025kdl}.

\section{Formation of shell-crossing singularities}
\label{sec:FormSCS}
While at the classical level SCS can be avoided by a suitable choice of initial data \cite{Lasky:2006hq}, in \cite{Fazzini:2023ova} it was shown that if an LQG-inspired effective model with a symmetric bounce is used to describe the gravitational collapse of dust spheres with inhomogenous profiles of compact support SCS arise despite the presence of quantum corrections. In what follows, we study whether this property also extends to more general model such as an LQG-inspired model with an asymmetric bounce \cite{Dapor:2017rwv} and the models by Bardeen \cite{2767662} and Hayward \cite{Hayward:2005gi}.
~\\
~\\
First, we start with reviewing the conditions under which SCS arise. For a given profile $M(x)$ and a given solution $R(t, x)$ the energy density of the dust reads
\begin{align}
    \rho(t, x) = \frac{M'(x)}{4 \pi R(t,x)^2 R'(t, x)}~.
\end{align}
A physical SCS in the dust region is then said to form if the conditions $R'(t, x) = 0$ (two different shells cross) and $M'(x) \neq 0$ (the crossing arises in the matter region) for the same shell $x$ hold as this leads to a divergence of the dust energy density and the curvature scalars. Physically, this is due to the crossing of two distinct matter layers travelling at different velocities. In that case the equations of motion break down and the LTB dynamics beyond the formation of a SCS loose their validity. Differently from the classical central singularity $R(t, x) = 0$ which can be avoided in effective models defined for instance in section \ref{sec:ReviewEffModels}, SCS are characterized by finite deviations of dust lines, even though the tidal forces may diverge. As a consequence, no infinite stretching or compression of matter occurs, see  \cite{Szekeres:1995gy} for the classical case and \cite{Fazzini:2025ysd} for the effective symmetric bounce scenario. This explains and justifies the effort to extend the dynamics beyond them.

A concrete realization of such an extension is to pass to the Painlevé-Gullstrand coordinates, in which case the formation of a SCS amounts to crossings of characteristics (multi-valued behavior of the dynamical fields). Then one looks for weak solutions satisfying the integral formulation of the original evolution equations \cite{Husain:2022gwp, Giesel:2023hys, Fazzini:2023ova, Fazzini:2025hsf}, where the multi-valued behavior of the LTB fields is replaced by an evolving gravitational shock. Recently, in \cite{Bobula:2026zlq}, a different approach for the symmetric bounce model was taken by treating the first SCS as a non-isolated thin shell and  using Darmois–Israel junction conditions. This makes it possible to control the signature of the thin shell surface—an aspect that is very difficult to handle when using weak solutions \cite{Fazzini:2025zrq}. In this work we will primarily discuss under which circumstances SCS form, further investigations towards spacetime extensions will be part of future works.
~\\
~\\
The following analysis is carried out in the marginally bound sector $\mathcal{E}(x) = 0$.
Whenever graphical illustrations are shown, we set the polymarization parameter $\alpha=1$ and assume $\gamma = 0.24$ following the results from black hole thermodynamics in the context of LQG \cite{Ashtekar:1997yu}. 

\subsection{Asymmetric bouncing model}\label{sec:A}

\begin{figure}
    \centering 
    \includegraphics[width=.5\linewidth]{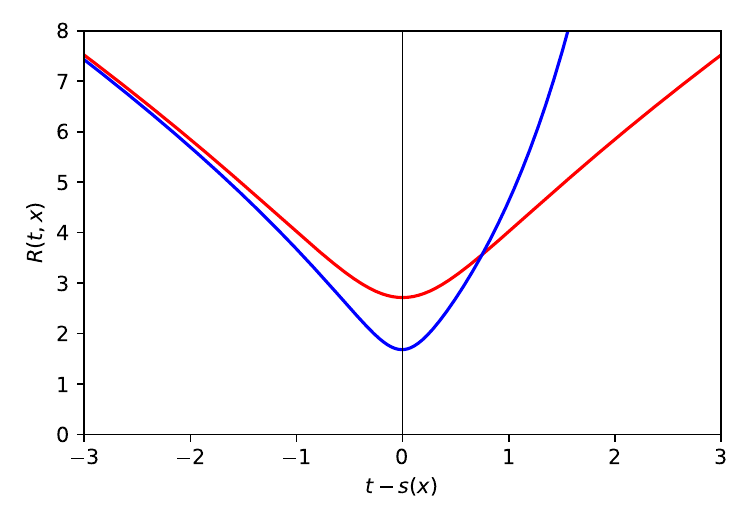}
    \caption{Plot of the areal radius as a function of $t - s(x)$ in the symmetric (red) and asymmetric (blue) bouncing model for a fixed mass $M(x) = 10$.} 
    \label{fig:Rbounce}
\end{figure}

\begin{figure}
    \centering \includegraphics[width=.5\linewidth]{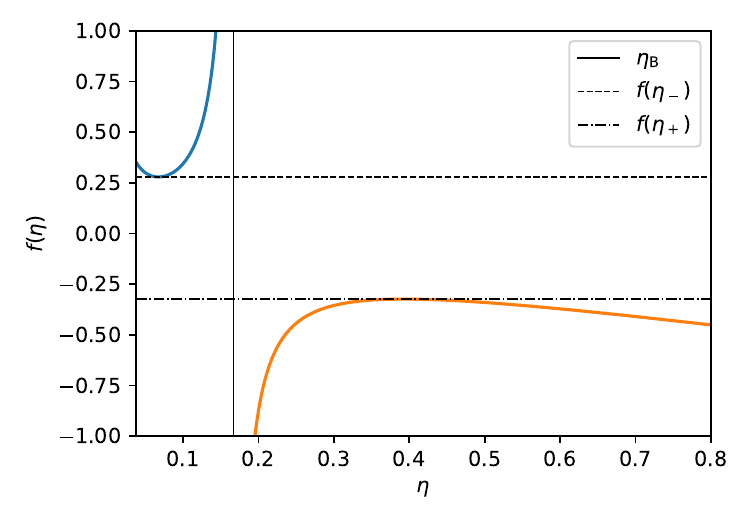}
    \caption{Plot of the function $f(\eta)$ defined in \eqref{eq:feta} in the pre- (orange) and post-bounce (blue) phase with the local extrema $f(\eta_\pm)$ indicated by the horizontal lines. } 
    \label{fig:f_eta}
\end{figure}

We begin our analysis with the asymmetric bouncing model which is inspired from Thiemann regularized LQC \cite{Yang:2009fp, Dapor:2017rwv, Li:2018opr}. The modified Friedmann equation provided in equation \eqref{eq:ModFRWAsymB} has an analytical solution given by
\begin{align}\label{eq:Rsol}
    R(t, x) = \sqrt[3]{\frac{2 G M(x) (4 \alpha_\Delta^2 \gamma^2 + 9 \eta^2)^2}{36 \eta^2 - 16 \alpha_\Delta^2 \gamma^4}},\hspace{.5cm} s(x) - t = \eta - \frac{2}{3} \alpha_\Delta (1 + \gamma^2) \tanh^{-1}\left(\frac{2 \alpha_\Delta \gamma^2}{3 \eta}\right),
\end{align}
where $s(x)$ is a time independent integration constant fixed by initial conditions and $\eta \in (2 \alpha_\Delta \gamma^2/3, \infty)$. The solution describes an asymmetric bouncing evolution for each dust shell with the minimal areal radius at the bounce $R_\mathrm{min} = 2 [\alpha_\Delta^2 \gamma^2 (\gamma^2 + 1) G M(x)]^{1/3}$ located at $\eta = \eta_\mathrm{B} \equiv 2/3 \gamma \alpha_\Delta (1 + 2 \gamma^2)^{1/2}$. Here the pre- and post-bounce phase are defined on the domains $\eta \ge \eta_\mathrm{B}$ and $2 \alpha_\Delta \gamma^2/3 < \eta \le \eta_\mathrm{B}$ respectively, so that $\eta$ decreases during the dynamics. In Fig. \ref{fig:Rbounce} the areal radius of a selected shell is plotted for a fixed mass $M(x) = 10$ as a function of $t - s(x)$ in order to illustrate the key differences compared to the standard symmetric bouncing model. In particular, one can notice that the pre-bounce phase has the same asymptotic (classical) limit in the two scenarios, while the post-bounce scenario is different, as expected from the seminal results in \cite{Assanioussi:2018hee, Assanioussi:2019iye} in the cosmological context. More precisely, while the symmetric case recovers a classical dynamics at late times, the asymmetric one consists in a DeSitter-like phase driven by an effective cosmological constant of Planckian order, therefore not settling down to a classical regime (see \cite{Assanioussi:2018hee, Assanioussi:2019iye} for a discussion on this).
~\\
~\\
In order to systematically investigate the formation of SCS within the asymmetric bouncing model, by using the solution in \eqref{eq:Rsol} we can rewrite the condition $R'(t, x) = 0$ as follows
\begin{align}\label{eq:feta}
    \frac{M(x) s'(x)}{M'(x)} = \frac{(4 \alpha_\Delta^2 \gamma^2 + 9 \eta^2)^2}{18 \eta [4 \alpha_\Delta^2 \gamma^2 (2 \gamma^2 + 1) - 9 \eta^2]} =: f(\eta).
\end{align}
To gain a clearer physical understanding of this condition, we express the left-hand side in terms of more intuitive physical quantities, as in the symmetric case \cite{Fazzini:2023ova}. To this end, we first refer to the analytical solution in \eqref{eq:Rsol} and define the scaling freedom of the areal radius such that $R(t_0, x) = x$ holds for a specific $\eta_0 > \eta_\mathrm{B}$ (pre-bounce phase), assuming a specific initial profile for the energy density. 
Then we can derive from the solution \eqref{eq:Rsol} computed at $t_0$ an expression for $s^\prime(x)$
\begin{align}\label{eq:8}
    s^\prime(x) = \frac{1 - x \frac{M'(x)}{3 M(x)}}{\eta_0 x} \frac{(4 \gamma^2 \alpha_\Delta^2 + 9 \eta_0^2)^2}{6 [9 \eta_0^2 - 4 \alpha_\Delta^2 \gamma^2 (1 + 2 \gamma^2)]},
\end{align}
which by construction is either entirely positive or negative, depending on the initial energy density profile under consideration as discussed in the following two subsections. In order to relate the condition for the formation of SCS in \eqref{eq:feta} to the choice of the intial density profile, we decompose it into an initial average energy density $\overline{\rho}_0$ and some deviation of it denoted as $\delta \rho_0$
\begin{align}\label{eq:EnergyDens}
    \rho(t_0, x) \equiv \rho_0 = \overline{\rho}_0(x) + \delta \rho_0(x),\hspace{.3cm}\overline{\rho}_0 = \frac{3 M(x)}{4\pi x^3}.
\end{align}
This allows us to express the left-hand side in \eqref{eq:feta} as 
\begin{align}
    \frac{M'(x)}{M(x)} = \frac{3 \rho_0}{x \overline{\rho}_0}.
\end{align}
The next step is to find a suitable expression for $s'(x)$ in \eqref{eq:8} in terms of the quantities mentioned above. In fact, solving the initial condition $R(t_0, x) = x$ for the respective parameter $\eta_0$ and using its relation to the temporal parameter $t$ we can finally rewrite the condition in \eqref{eq:feta} as
\begin{align}\label{eq:condrho}
    \frac{M(x) s'(x)}{M'(x)} = - \frac{\delta \rho_0 (\mathcal{A} + \sqrt{3})}{4 \rho_0 \mathcal{A} \sqrt{\pi G \overline{\rho}_0} \sqrt{- 16 \pi \alpha_\Delta^2 \gamma^2 G \overline{\rho}_0 + \sqrt{3} \mathcal{A} + \sqrt{3}}},
\end{align}
where $\mathcal{A} := [3 - 32 \pi \alpha_\Delta^2 \gamma^2 (\gamma^2 + 1)G \overline{\rho}_0]^{1/2}$.
Incidentally, when solving \eqref{eq:Rsol} for $\eta_0$ at the initial time, we obtain a total of four different roots, but only one of them guarantees the correct classical limit $\alpha_\Delta \to 0$ from \eqref{eq:condrho}, which is given by
\begin{align}
    \frac{M(x) s'(x)}{M'(x)} = - \frac{\delta \rho_0}{\rho_0} \frac{1}{\sqrt{8 \pi G \overline{\rho}_0}}~. 
\end{align}
A SCS forms in the dust region provided that we can find at least one $\eta$ in the given range for which the equation \eqref{eq:feta} is satisfied. An initial indication of whether SCS can be ruled out can be obtained by examining equations \eqref{eq:feta} and \eqref{eq:condrho}. Indeed, the right-hand side of \eqref{eq:feta} is positive (negative) in the post-bounce (pre-bounce) phase $\eta<\eta_B$ ($\eta>\eta_B$), while the left-hand side, given by \eqref{eq:condrho} is positive (negative) for decreasing (increasing) initial profiles, with $\delta \rho_0 <0 $ ($\delta \rho_0 >0 $). Therefore, no SCS can arise in the pre-bounce (post-bounce) phase for decreasing (increasing) initial profiles. A completely analogous situation as in the symmetric bouncing scenario investigated in \cite{Fazzini:2023ova}. In the next subsections we will deal with the remaining cases.
~\\
~\\
Instead of solving \eqref{eq:feta} for $\eta$, which makes the result highly dependent on the specific profile, we consider the extrema of the function $f(\eta)$ defined in \eqref{eq:feta} which are given by
\begin{align}
    \eta_{\pm} = \frac{2}{3}\alpha_\Delta \gamma \sqrt{3 \gamma^2 \pm \sqrt{9 \gamma^4 + 16\gamma^2 + 8} + 3},
\end{align}
where we dropped negative solutions as they do not belong to the domain of the areal radius. We realize that in general $\eta_+$ ($\eta_-$) is a local maximum (minimum) of the function $f(\eta)$ and that $\eta_+ > \eta_\mathrm{B}$, $\eta_- < \eta_\mathrm{B}$ for all positive $\gamma$ and $\alpha_\Delta$, see Fig. \ref{fig:f_eta} for a graphical illustration. In addition, we have that $\eta_- > 2 \alpha_\Delta \gamma^2 / 3$ only if $\gamma^2 (4 + 5 \gamma^2) < 1$ which is satisfied for $\gamma = 0.24$. We continue our analysis by distinguishing between initial decreasing and increasing profiles, since they lead to two qualitatively different scenarios.

\subsubsection{Initial decreasing profiles}
As indicated in the previous section, for initial decreasing profiles SCS can arise only in the post-bounce phase. Since $f(\eta_-)$ is a local minimum for the entire function, but a global minimum in the restricted domain $2 \alpha_\Delta \gamma^2 / 3<\eta<\eta_B$ (see Fig. \ref{fig:f_eta}), this means that a SCS forms after the bounce if
\begin{align}\label{eq:DPLB}
    \frac{M(x) s'(x)}{M'(x)} &\ge \frac{\gamma (\Gamma_-^2 + 1)}{3 \Gamma_- (2 \gamma^2 - \Gamma_-^2 + 1)} \alpha_\Delta 
    =0.28 \alpha_\Delta,
\end{align}
where we defined $\Gamma_- := [3 \gamma^2 - (9 \gamma^4 + 16\gamma^2 + 8)^{1/2} + 3]^{1/2}$. Even if quantitatively different, the situation is qualitatively the same as in the symmetric bouncing model, where the lower bound is given by $2 \gamma\alpha_\Delta / 3$ \cite{Fazzini:2023ova}.

From the lower bound obtained in \eqref{eq:DPLB}, the latest time at which a SCS can occur can be determined, since, according to the implicit solution in \eqref{eq:Rsol}, a decrease in the parameter $\eta$ starting from an initial point $\eta_0$ corresponds to a time evolution directed toward the future with respect to coordinate time $t$. Using \eqref{eq:Rsol} we explicitly obtain
\begin{align}
    t_\mathrm{latest} &= \frac{2}{3}\alpha_\Delta \left[(\gamma^2 + 1) \tanh^{-1}\left(\frac{\gamma}{\Gamma_-}\right) - \gamma \Gamma_- \right] + s(x)
    = 0.38 \alpha_\Delta + s(x).
\end{align}
Hence, after defining the time at which the bounce happens to be $t_\mathrm{B}(x)= s(x)$ as a reference point we find the following time interval in which SCS may form:
\begin{align}
    t_\mathrm{B} < t_\mathrm{SCS} \le t_\mathrm{B} + 0.38 \alpha_\Delta. 
\end{align}
Similar to the symmetric bouncing scenario \cite{Fazzini:2025hsf}, we can conclude that the time difference between the latest time and the bounce is of Planckian order.
~\\
~\\
By writing down the conditon for formation of SCS \eqref{eq:DPLB} in terms of the initial energy density \eqref{eq:condrho}, and defining its critical value at the bounce as \cite{Assanioussi:2018hee, tba}
\begin{align}
    \rho_\mathrm{c} = \frac{3}{32 \pi G \alpha_\Delta^2 \gamma^2 (1+\gamma^2)},
\end{align}
we can turn the inequality condition in \eqref{eq:DPLB} into the following form
\begin{equation}\label{eq:rhoc}
    \frac{|\delta \rho_0|}{\rho_0}\geq \frac{3k\sqrt{\frac{\overline{\rho}_0}{\rho_\mathrm{c}}}  \sqrt{1-\frac{\overline{\rho}_0}{\rho_\mathrm{c}}} \sqrt{1-\frac{ \overline{\rho}_0}{2(1+\gamma^2) \rho_\mathrm{c}} + \sqrt{1-\frac{\overline{\rho}_0}{\rho_\mathrm{c}}} }}{  \gamma \sqrt{2(\gamma^2+1)}\sqrt{1-\frac{\overline{\rho}_0}{\rho_\mathrm{c}} } + 1}~,
\end{equation}
where we used the fact that $\delta \rho_0 < 0$ for decreasing profiles and $k = 0.28$. As a result, for non-Planckian initial energy density profiles where $\rho_0 \ll 1$ and hence $\overline{\rho}_0 / \rho_\mathrm{c} \ll 1$, $\delta\rho_0 / \rho_\mathrm{c} \ll 1$ the lower bound becomes relatively small such that small inhomogeneities $|\delta \rho_0|$ are sufficient for SCS to develop. Note that in the case of exact homogeneity with $\delta \rho_0 = 0$ as in cosmological models, the condition \eqref{eq:rhoc} is never satisfied, and thus no SCS form. This result is in qualitative agreement with the symmetric bouncing scenario \cite{Fazzini:2023ova}.

\subsubsection{Initial increasing profiles}

As a second case we consider initial profiles with a global maximum at some $x>0$. In particular, we focus on the region where $\partial_x \rho_0 >0$. As discussed previously, the condition for the formation of SCS in \eqref{eq:feta} can be fulfilled only in the pre-bounce phase, and since $f(\eta_+)$ is a global maximum in the restricted temporal domain $\eta>\eta_B$ (see Fig. \ref{fig:f_eta}). In this time span we obtain the inequality
\begin{align}\label{eq:IPLB}
    \frac{M(x) s'(x)}{M'(x)} &\le -\frac{\gamma (\Gamma_+^2 + 1)}{3 \Gamma_+ (\Gamma_+^2 - 2 \gamma^2 - 1)} \alpha_\Delta 
    = -0.32 \alpha_\Delta
\end{align}
with $\Gamma_+ := [3 \gamma^2 + (9 \gamma^4 + 16\gamma^2 + 8)^{1/2} + 3]^{1/2}$. Likewise for initial decreasing profiles we can use the position of the maximum to determine the latest possible time for the formation of a SCS. Note that in this case they can form for arbitrary $\eta \ge \eta_+$ as $f(\eta)$ is monotonically decreasing in this range and $\lim_{\eta \to \infty} f(\eta) = -\infty$. Thus, we obtain the relation
\begin{align}
   t_0 \le t_\mathrm{SCS} < t_\mathrm{B} - 0.32 \alpha_\Delta ,
\end{align}
where $t_0$ denotes the inital time. In terms of the energy density we obtain exactly the same condition as in the previous case in \eqref{eq:rhoc} with the exception of some quantitative differences following from $\delta \rho_0 > 0$ and the different numerical value of the lower bound in \eqref{eq:IPLB}. This means that SCS will generally arise before the bounce in the increasing part of non-Planckian profiles. 

\subsection{Non-bouncing models: Bardeen \& Hayward}\label{sec:B}

\begin{figure}
    \centering \includegraphics[width=.5\linewidth]{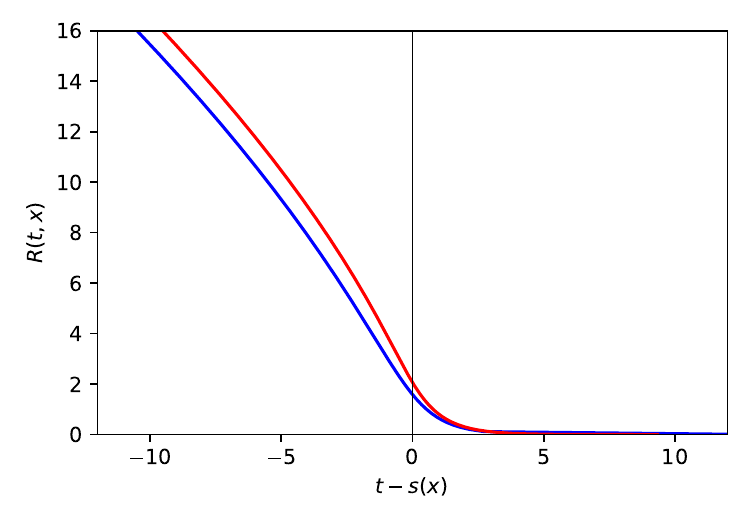}
    \caption{Plot of the areal radius as a function of $t - s(x)$ in the Hayward (red) and Bardeen (blue) model for a fixed mass $M(x) = 10$.}
    \label{fig:RBardeenHayward}
\end{figure}

In this section we apply the same procedure to dust collapsing models that are characterized by unbounded polymerization functions \cite{Giesel:2024mps}. More specifically, we will examine whether regular black hole models according to Bardeen \cite{2767662} and Hayward \cite{Hayward:2005gi}, when considered in the LTB context, exhibit similar features as bouncing models concerning the formation of SCS. Their implicit solutions in the marginally bound case read \cite{Giesel:2024mps}
\begin{widetext}
\begin{align}
    \text{Bardeen:}\hspace{.3cm}R(t, x) &= (2 G M(x))^{1/3} \sqrt{\eta^{4/3} - \alpha^{4/3}},\label{eq:RBardeen}\\ \hspace{.5cm} s(x) - t &= \frac{2}{3}\eta + \alpha \tan^{-1}\left(\frac{\eta^{1/3}}{\alpha^{1/3}}\right) - \alpha \mathrm{Re} \tanh^{-1}\left(\frac{\eta^{1/3}}{\alpha^{1/3}}\right) \nonumber \\
    \text{Hayward:}\hspace{.3cm}R(t, x) &= \left(\frac{2 G M(x) \alpha^2}{\sinh^2(\alpha \eta)}\right)^{1/3},\hspace{.5cm}s(x) - t = \frac{2}{3}\alpha \left(\coth(\alpha \eta) - \alpha\eta\right),\label{eq:RHayward}
\end{align}
\end{widetext}
where $\eta > \alpha$ and $\eta > 0$ respectively. As depicted in Fig. \ref{fig:RBardeenHayward} for $M(x) = 10$, both solutions describe an effective gravitational collapse without a bounce in which the resolution of the central singularity occurs due to the slow convergence of the areal radii to zero at $t \to \infty$.
Imposing the condition $R'(t, x) = 0$ as well as computing $s'(x)$ in terms of the initial energy density in \eqref{eq:EnergyDens} following the steps in the previous section leads to
\begin{align}
    \text{Bardeen:}\hspace{.3cm}\frac{M(x) s'(x)}{M'(x)} &= -\frac{\eta}{3} = -\frac{\delta \rho_0}{3 \rho_0} \left(\alpha^{4/3} + \left(\frac{3}{8 \pi G \overline \rho}_0\right)^{2/3}\right)^{3/4}, \\
    \text{Hayward:}\hspace{.3cm}\frac{M(x) s'(x)}{M'(x)} &= -\frac{1}{3} \alpha \coth (\alpha \eta) = -\frac{\delta \rho_0}{3 \rho_0} \sqrt{\alpha^2 + \frac{3}{8 \pi G \overline{\rho}_0}}.
\end{align}
Since both solutions for the areal radius are defined for $\eta > \alpha$ and $\eta > 0$ respectively, it immediately follows that the above equations are satisfied for $\delta \rho_0 > 0$ only, which implies initial profiles strictly increasing in at least one region of the domain, as in the classical collapse \cite{Hellaby:1985zz}. Taking furthermore into account that the functions in terms of $\eta$ are monotonic due to the absence of a bounce, one can conclude that SCS may form at arbitrary times, yet in accordance with classical collapse. Lastly, using the lower bounds for the parameter $\eta$ we obtain estimations of the form
\begin{align}
  \text{Bardeen:}\hspace{.3cm}&\frac{\delta \rho_0}{\rho_0} \ge \frac{1}{\left(1 + \left(\frac{3}{8 \pi G \alpha^2 \overline{\rho}_0}\right)^{2/3}\right)^{3/4}}~,\\
   \text{Hayward:}\hspace{.3cm}&\frac{\delta \rho_0}{\rho_0} \ge \frac{1}{\sqrt{1 + \frac{3}{8 \pi G \alpha^2 \overline{\rho}_0}}}~,
\end{align}
from which we can infer that small (positive) deviations $\delta \rho_0$ in the case of non-Planckian profiles are sufficient for the formation of SCS. 

\section{Conclusions}
\label{sec:Conclusion}
\noindent 
In the classical framework, when describing gravitational collapse within LTB models, we encounter two types of singularities: the central singularity and shell-crossing singularities (SCS), the latter of which can be avoided by fine-tuning the choice of  initial conditions \cite{Lasky:2006hq}. A common feature of the effective polymerized LTB models of the types considered in \cite{Giesel:2023tsj,Giesel:2023hys} is that the central singularity is resolved and replaced by regular dynamics for both homogeneous and inhomogeneous dust profiles. An interesting question is therefore whether or not a SCS can form in the effective LTB models. First studies can be found in in \cite{Fazzini:2023ova} where it was shown that for the symmetric bouncing model the case of the inhomogeneous stellar collapse leads unavoidably to SCS. In this work we extend the investigation to the asymmetric bouncing model derscribed by specific bounded poymerization functions, as well as to LTB models based on Bardeen \cite{2767662} and Hayward \cite{Hayward:2005gi} solutions which can be associated with unbounded polymerization functions. 
~\\
~\\
Our primary findings are that for an asymmetric bouncing model inspired by Loop Quantum Gravity (LQG), the inhomogeneous stellar collapse leads unavoidably to SCS. These form within approximately one Planck time after the bounce of the matter layers. Remarkably, this result is in strong qualitative agreement with the symmetric bouncing scenario discussed in \cite{Fazzini:2023ova}. This suggests that SCS may be a universal feature of LQG inspired bouncing models with bounded polymerization functions, at least for the class of quantum corrections encoded in these polymerization functions shared by these models. In contrast, in effective LTB models based on the solutions of Bardeen and Hayward, deviations from General Relativity lead to regular asymptotic dynamics rather than a bounce. In these cases, we have found that SCS are generally avoided for initially decreasing dust energy density profiles—the most physically relevant configurations for modeling stellar collapse. From a physical perspective, this is consistent: SCS in bouncing models arise primarily because adjacent shells bounce at different times. If the matter layers do not undergo a bounce, the formation of such singularities is suppressed.

Therefore, the formation of SCS serves as a distinguishing feature between regular models with bounce and regular models without a bounce here associated with bounded and unbounded respectively polymerization functions. A general proof of the above statement remains difficult to provide due to the enormous variety of regular effective gravitational collapse models. However, for bouncing scenarios, in which SCS necessarily develop, it is essential to establish physically grounded spacetime extensions, see for instance the work in \cite{Liu:2025fil,Husain:2025wrh,Bobula:2026zlq} for the symmetric bounding model. The implementation of this extension for the asymmetric model is left for future work. Another potential direction for future research is to conduct a similar study on the formation of SCS in models involving polymerized matter, in order to determine whether the presence of polymerized matter \cite{Hassan:2017cje} changes the situation or not.

\begin{acknowledgments}
E.R. thanks the Villigst foundation for financial support. K.G.\ is grateful for the hospitality of the Perimeter Institute, where part of this work was carried out. Research at the Perimeter Institute is supported in part by the Government of Canada through the Department of Innovation, Science and Economic Development, and by the Province of Ontario through the Ministry of Colleges and Universities. This work was supported by a grant from the Simons Foundation (Grant No.~1034867, Dittrich).
\end{acknowledgments}

\bibliographystyle{unsrtnat}
\bibliography{BibFormSCS}

\begin{thebibliography}{86}
\providecommand{\natexlab}[1]{#1}
\providecommand{\url}[1]{\texttt{#1}}
\expandafter\ifx\csname urlstyle\endcsname\relax
  \providecommand{\doi}[1]{doi: #1}\else
  \providecommand{\doi}{doi: \begingroup \urlstyle{rm}\Url}\fi

\bibitem[Abbott et~al.(2016)]{LIGOScientific:2016aoc}
B.~P. Abbott et~al.
\newblock {Observation of Gravitational Waves from a Binary Black Hole Merger}.
\newblock \emph{Phys. Rev. Lett.}, 116\penalty0 (6):\penalty0 061102, 2016.
\newblock \doi{10.1103/PhysRevLett.116.061102}.

\bibitem[{Webster} and {Murdin}(1972)]{1972Natur.235...37W}
B.~Louise {Webster} and Paul {Murdin}.
\newblock {Cygnus X-1-a Spectroscopic Binary with a Heavy Companion ?}
\newblock \emph{\nat}, 235\penalty0 (5332):\penalty0 37--38, January 1972.
\newblock \doi{10.1038/235037a0}.

\bibitem[Ghez et~al.(2008)]{Ghez:2008ms}
A.~M. Ghez et~al.
\newblock {Measuring Distance and Properties of the Milky Way's Central Supermassive Black Hole with Stellar Orbits}.
\newblock \emph{Astrophys. J.}, 689:\penalty0 1044--1062, 2008.
\newblock \doi{10.1086/592738}.

\bibitem[Akiyama et~al.(2019)]{EventHorizonTelescope:2019dse}
Kazunori Akiyama et~al.
\newblock {First M87 Event Horizon Telescope Results. I. The Shadow of the Supermassive Black Hole}.
\newblock \emph{Astrophys. J. Lett.}, 875:\penalty0 L1, 2019.
\newblock \doi{10.3847/2041-8213/ab0ec7}.

\bibitem[Oppenheimer and Snyder(1939)]{Oppenheimer:1939ue}
J.~R. Oppenheimer and H.~Snyder.
\newblock {On Continued gravitational contraction}.
\newblock \emph{Phys. Rev.}, 56:\penalty0 455--459, 1939.
\newblock \doi{10.1103/PhysRev.56.455}.

\bibitem[Lemaitre(1933)]{Lemaitre:1933gd}
G.~Lemaitre.
\newblock {The expanding universe}.
\newblock \emph{Annales Soc. Sci. Bruxelles A}, 53:\penalty0 51--85, 1933.
\newblock \doi{10.1023/A:1018855621348}.

\bibitem[Tolman(1934)]{Tolman:1934za}
Richard~C. Tolman.
\newblock {Effect of imhomogeneity on cosmological models}.
\newblock \emph{Proc. Nat. Acad. Sci.}, 20:\penalty0 169--176, 1934.
\newblock \doi{10.1073/pnas.20.3.169}.

\bibitem[Bondi(1947)]{Bondi:1947fta}
H.~Bondi.
\newblock {Spherically symmetrical models in general relativity}.
\newblock \emph{Mon. Not. Roy. Astron. Soc.}, 107:\penalty0 410--425, 1947.
\newblock \doi{10.1093/mnras/107.5-6.410}.

\bibitem[Penrose(1965)]{Penrose:1964wq}
Roger Penrose.
\newblock {Gravitational collapse and space-time singularities}.
\newblock \emph{Phys. Rev. Lett.}, 14:\penalty0 57--59, 1965.
\newblock \doi{10.1103/PhysRevLett.14.57}.

\bibitem[Joshi et~al.(2024)Joshi, Dey, Joshi, and Tank]{Joshi:2024djy}
Ashok~B. Joshi, Dipanjan Dey, Pankaj~S. Joshi, and Vivekkumar~R. Tank.
\newblock {Tidal forces in collapsing compact objects}.
\newblock \emph{Phys. Rev. D}, 110\penalty0 (12):\penalty0 124066, 2024.
\newblock \doi{10.1103/PhysRevD.110.124066}.

\bibitem[Szekeres and Lun(1999)]{Szekeres:1995gy}
Peter Szekeres and Anthony Lun.
\newblock {What is a shell crossing singularity?}
\newblock \emph{J. Austral. Math. Soc. B}, 41:\penalty0 167--179, 1999.
\newblock \doi{10.1017/S0334270000011140}.

\bibitem[Müller~zum Hagen et~al.(1974)Müller~zum Hagen, Yodzis, and Seifert]{hagendas1974}
H.~Müller~zum Hagen, P.~Yodzis, and H.J. Seifert.
\newblock {On the occurrence of naked singularities in general relativity. II.}
\newblock \emph{Commun. Math. Phys.}, 37:\penalty0 29--40, 1974.
\newblock \doi{10.1007/BF01646032}.

\bibitem[Hellaby and Lake(1985)]{Hellaby:1985zz}
C.~Hellaby and K.~Lake.
\newblock {Shell crossings and the Tolman model}.
\newblock \emph{Astrophys. J.}, 290:\penalty0 381, 1985.
\newblock \doi{10.1086/162995}.

\bibitem[Nolan(2003)]{Nolan:2003wp}
Brien~C. Nolan.
\newblock {Dynamical extensions for shell crossing singularities}.
\newblock \emph{Class. Quant. Grav.}, 20:\penalty0 575--586, 2003.
\newblock \doi{10.1088/0264-9381/20/4/302}.

\bibitem[Husain and Mehmood(2025)]{Husain:2025wrh}
Viqar Husain and Hassan Mehmood.
\newblock {Shock waves in classical dust collapse}.
\newblock \emph{Phys. Rev. Res.}, 7\penalty0 (3):\penalty0 033215, 2025.
\newblock \doi{10.1103/b317-86qs}.

\bibitem[Maeda and Sato(1983)]{Maeda1983}
Kei-ichi Maeda and Humitaka Sato.
\newblock Expansion of a thin shell around a void in expanding universe.
\newblock \emph{Progress of Theoretical Physics}, 70\penalty0 (3):\penalty0 772--782, 1983.
\newblock \doi{10.1143/PTP.70.772}.

\bibitem[Fazzini and Mehmood(2025)]{Fazzini:2025zrq}
Francesco Fazzini and Hassan Mehmood.
\newblock {Weak solutions in Einstein theory and beyond}.
\newblock \emph{Phys. Rev. D}, 112\penalty0 (6):\penalty0 064064, 2025.
\newblock \doi{10.1103/6mrq-f9qc}.

\bibitem[Lewandowski et~al.(2023)Lewandowski, Ma, Yang, and Zhang]{Lewandowski:2022zce}
Jerzy Lewandowski, Yongge Ma, Jinsong Yang, and Cong Zhang.
\newblock {Quantum Oppenheimer-Snyder and Swiss Cheese Models}.
\newblock \emph{Phys. Rev. Lett.}, 130\penalty0 (10):\penalty0 101501, 2023.
\newblock \doi{10.1103/PhysRevLett.130.101501}.

\bibitem[Husain et~al.(2022{\natexlab{a}})Husain, Kelly, Santacruz, and Wilson-Ewing]{Husain:2022gwp}
Viqar Husain, Jarod~George Kelly, Robert Santacruz, and Edward Wilson-Ewing.
\newblock {Fate of quantum black holes}.
\newblock \emph{Phys. Rev. D}, 106\penalty0 (2):\penalty0 024014, 2022{\natexlab{a}}.
\newblock \doi{10.1103/PhysRevD.106.024014}.

\bibitem[Bobula and Paw{\l}owski(2023)]{Bobula:2023kbo}
Micha{\l} Bobula and Tomasz Paw{\l}owski.
\newblock {Rainbow Oppenheimer-Snyder collapse and the entanglement entropy production}.
\newblock \emph{Phys. Rev. D}, 108\penalty0 (2):\penalty0 026016, 2023.
\newblock \doi{10.1103/PhysRevD.108.026016}.

\bibitem[Alonso-Bardaji and Brizuela(2024)]{Alonso-Bardaji:2023qgu}
Asier Alonso-Bardaji and David Brizuela.
\newblock {Nonsingular collapse of a spherical dust cloud}.
\newblock \emph{Phys. Rev. D}, 109\penalty0 (6):\penalty0 064023, 2024.
\newblock \doi{10.1103/PhysRevD.109.064023}.

\bibitem[Bojowald et~al.(2025)Bojowald, Duque, and Hartmann]{Bojowald:2024ium}
Martin Bojowald, Erick~I. Duque, and Dennis Hartmann.
\newblock {Covariant Lema{\^\i}tre-Tolman-Bondi collapse in models of loop quantum gravity}.
\newblock \emph{Phys. Rev. D}, 111\penalty0 (6):\penalty0 064002, 2025.
\newblock \doi{10.1103/PhysRevD.111.064002}.

\bibitem[Han et~al.(2023)Han, Rovelli, and Soltani]{Han:2023wxg}
Muxin Han, Carlo Rovelli, and Farshid Soltani.
\newblock {Geometry of the black-to-white hole transition within a single asymptotic region}.
\newblock \emph{Phys. Rev. D}, 107\penalty0 (6):\penalty0 064011, 2023.
\newblock \doi{10.1103/PhysRevD.107.064011}.

\bibitem[Cafaro and Lewandowski(2024)]{Cafaro:2024vrw}
Luca Cafaro and Jerzy Lewandowski.
\newblock {Status of Birkhoff{\textquoteright}s theorem in the polymerized semiclassical regime of loop quantum gravity}.
\newblock \emph{Phys. Rev. D}, 110\penalty0 (2):\penalty0 024072, 2024.
\newblock \doi{10.1103/PhysRevD.110.024072}.

\bibitem[Liu and Qu(2025)]{Liu:2025fil}
Hongguang Liu and Dongxue Qu.
\newblock {Quantum induced shock dynamics in gravitational collapse: insights from effective models and numerical frameworks}.
\newblock 4 2025.

\bibitem[Ashtekar and Bojowald(2006)]{Ashtekar:2005qt}
Abhay Ashtekar and Martin Bojowald.
\newblock {Quantum geometry and the Schwarzschild singularity}.
\newblock \emph{Class. Quant. Grav.}, 23:\penalty0 391--411, 2006.
\newblock \doi{10.1088/0264-9381/23/2/008}.

\bibitem[Modesto(2006)]{Modesto:2005zm}
Leonardo Modesto.
\newblock {Loop quantum black hole}.
\newblock \emph{Class. Quant. Grav.}, 23:\penalty0 5587--5602, 2006.
\newblock \doi{10.1088/0264-9381/23/18/006}.

\bibitem[Boehmer and Vandersloot(2007)]{Boehmer:2007ket}
Christian~G. Boehmer and Kevin Vandersloot.
\newblock {Loop Quantum Dynamics of the Schwarzschild Interior}.
\newblock \emph{Phys. Rev. D}, 76:\penalty0 104030, 2007.
\newblock \doi{10.1103/PhysRevD.76.104030}.

\bibitem[Chiou et~al.(2012)Chiou, Ni, and Tang]{Chiou:2012pg}
Dah-Wei Chiou, Wei-Tou Ni, and Alf Tang.
\newblock {Loop quantization of spherically symmetric midisuperspaces and loop quantum geometry of the maximally extended Schwarzschild spacetime}.
\newblock 12 2012.

\bibitem[Gambini et~al.(2014)Gambini, Olmedo, and Pullin]{Gambini:2013hna}
Rodolfo Gambini, Javier Olmedo, and Jorge Pullin.
\newblock {Quantum black holes in Loop Quantum Gravity}.
\newblock \emph{Class. Quant. Grav.}, 31:\penalty0 095009, 2014.
\newblock \doi{10.1088/0264-9381/31/9/095009}.

\bibitem[Brahma(2015)]{Brahma:2014gca}
Suddhasattwa Brahma.
\newblock {Spherically symmetric canonical quantum gravity}.
\newblock \emph{Phys. Rev. D}, 91\penalty0 (12):\penalty0 124003, 2015.
\newblock \doi{10.1103/PhysRevD.91.124003}.

\bibitem[Dadhich et~al.(2015)Dadhich, Joe, and Singh]{Dadhich:2015ora}
Naresh Dadhich, Anton Joe, and Parampreet Singh.
\newblock {Emergence of the product of constant curvature spaces in loop quantum cosmology}.
\newblock \emph{Class. Quant. Grav.}, 32\penalty0 (18):\penalty0 185006, 2015.
\newblock \doi{10.1088/0264-9381/32/18/185006}.

\bibitem[Tibrewala(2014)]{Tibrewala:2013kba}
Rakesh Tibrewala.
\newblock {Inhomogeneities, loop quantum gravity corrections, constraint algebra and general covariance}.
\newblock \emph{Class. Quant. Grav.}, 31:\penalty0 055010, 2014.
\newblock \doi{10.1088/0264-9381/31/5/055010}.

\bibitem[Ben~Achour et~al.(2018{\natexlab{a}})Ben~Achour, Lamy, Liu, and Noui]{BenAchour:2017ivq}
Jibril Ben~Achour, Frederic Lamy, Hongguang Liu, and Karim Noui.
\newblock {Non-singular black holes and the Limiting Curvature Mechanism: A Hamiltonian perspective}.
\newblock \emph{JCAP}, 05:\penalty0 072, 2018{\natexlab{a}}.
\newblock \doi{10.1088/1475-7516/2018/05/072}.

\bibitem[Yonika et~al.(2018)Yonika, Khanna, and Singh]{Yonika:2017qgo}
Alec Yonika, Gaurav Khanna, and Parampreet Singh.
\newblock {Von-Neumann Stability and Singularity Resolution in Loop Quantized Schwarzschild Black Hole}.
\newblock \emph{Class. Quant. Grav.}, 35\penalty0 (4):\penalty0 045007, 2018.
\newblock \doi{10.1088/1361-6382/aaa18d}.

\bibitem[D'Ambrosio et~al.(2021)D'Ambrosio, Christodoulou, Martin-Dussaud, Rovelli, and Soltani]{DAmbrosio:2020mut}
Fabio D'Ambrosio, Marios Christodoulou, Pierre Martin-Dussaud, Carlo Rovelli, and Farshid Soltani.
\newblock {End of a black hole{\textquoteright}s evaporation}.
\newblock \emph{Phys. Rev. D}, 103\penalty0 (10):\penalty0 106014, 2021.
\newblock \doi{10.1103/PhysRevD.103.106014}.

\bibitem[Olmedo et~al.(2017)Olmedo, Saini, and Singh]{Olmedo:2017lvt}
Javier Olmedo, Sahil Saini, and Parampreet Singh.
\newblock {From black holes to white holes: a quantum gravitational, symmetric bounce}.
\newblock \emph{Class. Quant. Grav.}, 34\penalty0 (22):\penalty0 225011, 2017.
\newblock \doi{10.1088/1361-6382/aa8da8}.

\bibitem[Ashtekar et~al.(2018{\natexlab{a}})Ashtekar, Olmedo, and Singh]{Ashtekar:2018lag}
Abhay Ashtekar, Javier Olmedo, and Parampreet Singh.
\newblock {Quantum Transfiguration of Kruskal Black Holes}.
\newblock \emph{Phys. Rev. Lett.}, 121\penalty0 (24):\penalty0 241301, 2018{\natexlab{a}}.
\newblock \doi{10.1103/PhysRevLett.121.241301}.

\bibitem[Ashtekar et~al.(2018{\natexlab{b}})Ashtekar, Olmedo, and Singh]{Ashtekar:2018cay}
Abhay Ashtekar, Javier Olmedo, and Parampreet Singh.
\newblock {Quantum extension of the Kruskal spacetime}.
\newblock \emph{Phys. Rev. D}, 98\penalty0 (12):\penalty0 126003, 2018{\natexlab{b}}.
\newblock \doi{10.1103/PhysRevD.98.126003}.

\bibitem[Bojowald et~al.(2018)Bojowald, Brahma, and Yeom]{Bojowald:2018xxu}
Martin Bojowald, Suddhasattwa Brahma, and Dong-han Yeom.
\newblock {Effective line elements and black-hole models in canonical loop quantum gravity}.
\newblock \emph{Phys. Rev. D}, 98\penalty0 (4):\penalty0 046015, 2018.
\newblock \doi{10.1103/PhysRevD.98.046015}.

\bibitem[Ben~Achour et~al.(2018{\natexlab{b}})Ben~Achour, Lamy, Liu, and Noui]{BenAchour:2018khr}
Jibril Ben~Achour, Fr{\'e}d{\'e}ric Lamy, Hongguang Liu, and Karim Noui.
\newblock {Polymer Schwarzschild black hole: An effective metric}.
\newblock \emph{EPL}, 123\penalty0 (2):\penalty0 20006, 2018{\natexlab{b}}.
\newblock \doi{10.1209/0295-5075/123/20006}.

\bibitem[Bodendorfer et~al.(2019)Bodendorfer, Mele, and M{\"u}nch]{Bodendorfer:2019cyv}
Norbert Bodendorfer, Fabio~M. Mele, and Johannes M{\"u}nch.
\newblock {Effective Quantum Extended Spacetime of Polymer Schwarzschild Black Hole}.
\newblock \emph{Class. Quant. Grav.}, 36\penalty0 (19):\penalty0 195015, 2019.
\newblock \doi{10.1088/1361-6382/ab3f16}.

\bibitem[Alesci et~al.(2019)Alesci, Bahrami, and Pranzetti]{Alesci:2019pbs}
Emanuele Alesci, Sina Bahrami, and Daniele Pranzetti.
\newblock {Quantum gravity predictions for black hole interior geometry}.
\newblock \emph{Phys. Lett. B}, 797:\penalty0 134908, 2019.
\newblock \doi{10.1016/j.physletb.2019.134908}.

\bibitem[Assanioussi et~al.(2020)Assanioussi, Dapor, and Liegener]{Assanioussi:2019twp}
Mehdi Assanioussi, Andrea Dapor, and Klaus Liegener.
\newblock {Perspectives on the dynamics in a loop quantum gravity effective description of black hole interiors}.
\newblock \emph{Phys. Rev. D}, 101\penalty0 (2):\penalty0 026002, 2020.
\newblock \doi{10.1103/PhysRevD.101.026002}.

\bibitem[Benitez et~al.(2020)Benitez, Gambini, Lehner, Liebling, and Pullin]{Benitez:2020szx}
Florencia Benitez, Rodolfo Gambini, Luis Lehner, Steve Liebling, and Jorge Pullin.
\newblock {Critical collapse of a scalar field in semiclassical loop quantum gravity}.
\newblock \emph{Phys. Rev. Lett.}, 124\penalty0 (7):\penalty0 071301, 2020.
\newblock \doi{10.1103/PhysRevLett.124.071301}.

\bibitem[Gan et~al.(2020)Gan, Santos, Shu, and Wang]{Gan:2020dkb}
Wen-Cong Gan, Nilton~O. Santos, Fu-Wen Shu, and Anzhong Wang.
\newblock {Properties of the spherically symmetric polymer black holes}.
\newblock \emph{Phys. Rev. D}, 102:\penalty0 124030, 2020.
\newblock \doi{10.1103/PhysRevD.102.124030}.

\bibitem[Gambini et~al.(2021)Gambini, Olmedo, and Pullin]{Gambini:2020qhx}
Rodolfo Gambini, Javier Olmedo, and Jorge Pullin.
\newblock {Loop Quantum Black Hole Extensions Within the Improved Dynamics}.
\newblock \emph{Front. Astron. Space Sci.}, 8:\penalty0 74, 2021.
\newblock \doi{10.3389/fspas.2021.647241}.

\bibitem[Husain et~al.(2022{\natexlab{b}})Husain, Kelly, Santacruz, and Wilson-Ewing]{Husain:2021ojz}
Viqar Husain, Jarod~George Kelly, Robert Santacruz, and Edward Wilson-Ewing.
\newblock {Quantum Gravity of Dust Collapse: Shock Waves from Black Holes}.
\newblock \emph{Phys. Rev. Lett.}, 128\penalty0 (12):\penalty0 121301, 2022{\natexlab{b}}.
\newblock \doi{10.1103/PhysRevLett.128.121301}.

\bibitem[Li and Singh(2021)]{Li:2021snn}
Bao-Fei Li and Parampreet Singh.
\newblock {Does the Loop Quantum $\mu_o$ Scheme Permit Black Hole Formation?}
\newblock \emph{Universe}, 7\penalty0 (11):\penalty0 406, 2021.
\newblock \doi{10.3390/universe7110406}.

\bibitem[Gan et~al.(2022)Gan, Ongole, Alesci, An, Shu, and Wang]{Gan:2022mle}
Wen-Cong Gan, Geeth Ongole, Emanuele Alesci, Yang An, Fu-Wen Shu, and Anzhong Wang.
\newblock {Understanding quantum black holes from quantum reduced loop gravity}.
\newblock \emph{Phys. Rev. D}, 106\penalty0 (12):\penalty0 126013, 2022.
\newblock \doi{10.1103/PhysRevD.106.126013}.

\bibitem[Kelly et~al.(2020)Kelly, Santacruz, and Wilson-Ewing]{Kelly:2020uwj}
Jarod~George Kelly, Robert Santacruz, and Edward Wilson-Ewing.
\newblock {Effective loop quantum gravity framework for vacuum spherically symmetric spacetimes}.
\newblock \emph{Phys. Rev. D}, 102\penalty0 (10):\penalty0 106024, 2020.
\newblock \doi{10.1103/PhysRevD.102.106024}.

\bibitem[Gambini et~al.(2020)Gambini, Olmedo, and Pullin]{Gambini:2020nsf}
R.~Gambini, J.~Olmedo, and J.~Pullin.
\newblock {Spherically symmetric loop quantum gravity: analysis of improved dynamics}.
\newblock \emph{Class. Quant. Grav.}, 37\penalty0 (20):\penalty0 205012, 2020.
\newblock \doi{10.1088/1361-6382/aba842}.

\bibitem[Han and Liu(2022)]{Han:2020uhb}
Muxin Han and Hongguang Liu.
\newblock {Improved effective dynamics of loop-quantum-gravity black hole and Nariai limit}.
\newblock \emph{Class. Quant. Grav.}, 39\penalty0 (3):\penalty0 035011, 2022.
\newblock \doi{10.1088/1361-6382/ac44a0}.

\bibitem[Zhang(2021)]{Zhang:2021xoa}
Cong Zhang.
\newblock {Reduced phase space quantization of black holes: Path integrals and effective dynamics}.
\newblock \emph{Phys. Rev. D}, 104\penalty0 (12):\penalty0 126003, 2021.
\newblock \doi{10.1103/PhysRevD.104.126003}.

\bibitem[M{\"u}nch et~al.(2023)M{\"u}nch, Perez, Speziale, and Viollet]{Munch:2022teq}
Johannes M{\"u}nch, Alejandro Perez, Simone Speziale, and Sami Viollet.
\newblock {Generic features of a polymer quantum black hole}.
\newblock \emph{Class. Quant. Grav.}, 40\penalty0 (13):\penalty0 135003, 2023.
\newblock \doi{10.1088/1361-6382/accccd}.

\bibitem[Giesel et~al.(2021)Giesel, Li, and Singh]{Giesel:2021dug}
Kristina Giesel, Bao-Fei Li, and Parampreet Singh.
\newblock {Nonsingular quantum gravitational dynamics of an Lema{\^\i}tre-Tolman-Bondi dust shell model: The role of quantization prescriptions}.
\newblock \emph{Phys. Rev. D}, 104\penalty0 (10):\penalty0 106017, 2021.
\newblock \doi{10.1103/PhysRevD.104.106017}.

\bibitem[Giesel et~al.(2023)Giesel, Han, Li, Liu, and Singh]{Giesel:2022rxi}
Kristina Giesel, Muxin Han, Bao-Fei Li, Hongguang Liu, and Parampreet Singh.
\newblock {Spherical symmetric gravitational collapse of a dust cloud: Polymerized dynamics in reduced phase space}.
\newblock \emph{Phys. Rev. D}, 107\penalty0 (4):\penalty0 044047, 2023.
\newblock \doi{10.1103/PhysRevD.107.044047}.

\bibitem[Han and Liu(2024)]{Han:2022rsx}
Muxin Han and Hongguang Liu.
\newblock {Covariant {\ensuremath{\mu}}{\textasciimacron}-scheme effective dynamics, mimetic gravity, and nonsingular black holes: Applications to spherically symmetric quantum gravity}.
\newblock \emph{Phys. Rev. D}, 109\penalty0 (8):\penalty0 084033, 2024.
\newblock \doi{10.1103/PhysRevD.109.084033}.

\bibitem[Giesel et~al.(2024{\natexlab{a}})Giesel, Liu, Rullit, Singh, and Weigl]{Giesel:2023tsj}
Kristina Giesel, Hongguang Liu, Eric Rullit, Parampreet Singh, and Stefan~Andreas Weigl.
\newblock {Embedding generalized Lema{\^\i}tre-Tolman-Bondi models in polymerized spherically symmetric spacetimes}.
\newblock \emph{Phys. Rev. D}, 110\penalty0 (10):\penalty0 104017, 2024{\natexlab{a}}.
\newblock \doi{10.1103/PhysRevD.110.104017}.

\bibitem[Giesel et~al.(2025)Giesel, Liu, Singh, and Weigl]{Giesel:2024mps}
Kristina Giesel, Hongguang Liu, Parampreet Singh, and Stefan~Andreas Weigl.
\newblock {Regular black holes and their relationship to polymerized models and mimetic gravity}.
\newblock \emph{Phys. Rev. D}, 111\penalty0 (6):\penalty0 064064, 2025.
\newblock \doi{10.1103/PhysRevD.111.064064}.

\bibitem[Taveras(2008)]{Taveras:2008ke}
Victor Taveras.
\newblock {Corrections to the Friedmann Equations from LQG for a Universe with a Free Scalar Field}.
\newblock \emph{Phys. Rev. D}, 78:\penalty0 064072, 2008.
\newblock \doi{10.1103/PhysRevD.78.064072}.

\bibitem[Giesel and Thiemann(2010)]{Giesel:2007wn}
K.~Giesel and T.~Thiemann.
\newblock {Algebraic quantum gravity (AQG). IV. Reduced phase space quantisation of loop quantum gravity}.
\newblock \emph{Class. Quant. Grav.}, 27:\penalty0 175009, 2010.
\newblock \doi{10.1088/0264-9381/27/17/175009}.

\bibitem[Dapor and Liegener(2018)]{Dapor:2017rwv}
Andrea Dapor and Klaus Liegener.
\newblock {Cosmological Effective Hamiltonian from full Loop Quantum Gravity Dynamics}.
\newblock \emph{Phys. Lett. B}, 785:\penalty0 506--510, 2018.
\newblock \doi{10.1016/j.physletb.2018.09.005}.

\bibitem[Giesel et~al.(2026)Giesel, Liu, and Rullit]{Giesel:2026pjj}
Kristina Giesel, Hongguang Liu, and Eric Rullit.
\newblock {Investigation of the gravitational dust collapse of the LQG-inspired effective asymmetric bounce model}.
\newblock 2 2026.

\bibitem[Bardeen()]{2767662}
James Bardeen.
\newblock {Nonsingular general relativistic gravitational collapse}.

\bibitem[Hayward(2006)]{Hayward:2005gi}
Sean~A. Hayward.
\newblock {Formation and evaporation of regular black holes}.
\newblock \emph{Phys. Rev. Lett.}, 96:\penalty0 031103, 2006.
\newblock \doi{10.1103/PhysRevLett.96.031103}.

\bibitem[Fazzini et~al.(2024)Fazzini, Husain, and Wilson-Ewing]{Fazzini:2023ova}
Francesco Fazzini, Viqar Husain, and Edward Wilson-Ewing.
\newblock {Shell-crossings and shock formation during gravitational collapse in effective loop quantum gravity}.
\newblock \emph{Phys. Rev. D}, 109\penalty0 (8):\penalty0 084052, 2024.
\newblock \doi{10.1103/PhysRevD.109.084052}.

\bibitem[Giesel et~al.(2024{\natexlab{b}})Giesel, Liu, Singh, and Weigl]{Giesel:2023hys}
Kristina Giesel, Hongguang Liu, Parampreet Singh, and Stefan~Andreas Weigl.
\newblock {Generalized analysis of a dust collapse in effective loop quantum gravity: Fate of shocks and covariance}.
\newblock \emph{Phys. Rev. D}, 110\penalty0 (10):\penalty0 104016, 2024{\natexlab{b}}.
\newblock \doi{10.1103/PhysRevD.110.104016}.

\bibitem[Bojowald et~al.(2008)Bojowald, Harada, and Tibrewala]{Bojowald:2008ja}
Martin Bojowald, Tomohiro Harada, and Rakesh Tibrewala.
\newblock {Lemaitre-Tolman-Bondi collapse from the perspective of loop quantum gravity}.
\newblock \emph{Phys. Rev. D}, 78:\penalty0 064057, 2008.
\newblock \doi{10.1103/PhysRevD.78.064057}.

\bibitem[Bojowald et~al.(2009)Bojowald, Reyes, and Tibrewala]{Bojowald:2009ih}
Martin Bojowald, Juan~D. Reyes, and Rakesh Tibrewala.
\newblock {Non-marginal LTB-like models with inverse triad corrections from loop quantum gravity}.
\newblock \emph{Phys. Rev. D}, 80:\penalty0 084002, 2009.
\newblock \doi{10.1103/PhysRevD.80.084002}.

\bibitem[Alonso-Bardaji et~al.(2022)Alonso-Bardaji, Brizuela, and Vera]{Alonso-Bardaji:2021yls}
Asier Alonso-Bardaji, David Brizuela, and Ra{\"u}l Vera.
\newblock {An effective model for the quantum Schwarzschild black hole}.
\newblock \emph{Phys. Lett. B}, 829:\penalty0 137075, 2022.
\newblock \doi{10.1016/j.physletb.2022.137075}.

\bibitem[Rovelli and Smolin(1995)]{Rovelli:1994ge}
Carlo Rovelli and Lee Smolin.
\newblock {Discreteness of area and volume in quantum gravity}.
\newblock \emph{Nucl. Phys. B}, 442:\penalty0 593--622, 1995.
\newblock \doi{10.1016/0550-3213(95)00150-Q}.
\newblock [Erratum: Nucl.Phys.B 456, 753--754 (1995)].

\bibitem[Ashtekar and Lewandowski(1997)]{Ashtekar:1996eg}
Abhay Ashtekar and Jerzy Lewandowski.
\newblock {Quantum theory of geometry. 1: Area operators}.
\newblock \emph{Class. Quant. Grav.}, 14:\penalty0 A55--A82, 1997.
\newblock \doi{10.1088/0264-9381/14/1A/006}.

\bibitem[Yang et~al.(2009)Yang, Ding, and Ma]{Yang:2009fp}
Jinsong Yang, You Ding, and Yongge Ma.
\newblock {Alternative quantization of the Hamiltonian in loop quantum cosmology II: Including the Lorentz term}.
\newblock \emph{Phys. Lett. B}, 682:\penalty0 1--7, 2009.
\newblock \doi{10.1016/j.physletb.2009.10.072}.

\bibitem[Bobula and Paw{\l}owski(2025)]{Bobula:2024chr}
Micha{\l} Bobula and Tomasz Paw{\l}owski.
\newblock {Causal structure of nonhomogeneous dust collapse in effective loop quantum gravity}.
\newblock \emph{Phys. Rev. D}, 112\penalty0 (6):\penalty0 064056, 2025.
\newblock \doi{10.1103/z5sz-mcl5}.

\bibitem[Giesel and Liu(2025)]{Giesel:2025kdl}
Kristina Giesel and Hongguang Liu.
\newblock {From Principles to Effective Models: A Constructive Framework for Effective Covariant Actions with a Unique Vacuum Solution}.
\newblock 12 2025.

\bibitem[Lasky et~al.(2006)Lasky, Lun, and Burston]{Lasky:2006hq}
Paul~D. Lasky, Anthony W.~C. Lun, and Raymond~B. Burston.
\newblock {Initial value formalism for dust collapse}.
\newblock 6 2006.

\bibitem[Fazzini(2025{\natexlab{a}})]{Fazzini:2025ysd}
Francesco Fazzini.
\newblock {Gentle spaghettification in effective LQG dust collapse}.
\newblock \emph{Phys. Rev. D}, 112\penalty0 (2):\penalty0 026029, 2025{\natexlab{a}}.
\newblock \doi{10.1103/f7q9-grqb}.

\bibitem[Fazzini(2025{\natexlab{b}})]{Fazzini:2025hsf}
Francesco Fazzini.
\newblock {Non-uniqueness of shockwave evolution in a loop quantum gravity inspired model}.
\newblock \emph{Phys. Scripta}, 100\penalty0 (11):\penalty0 115220, 2025{\natexlab{b}}.
\newblock \doi{10.1088/1402-4896/ae1be4}.

\bibitem[Bobula and Fazzini(2026)]{Bobula:2026zlq}
Micha{\l} Bobula and Francesco Fazzini.
\newblock {Quantum gravitational stellar evolution beyond shell-crossing singularities}.
\newblock 1 2026.

\bibitem[Ashtekar et~al.(1998)Ashtekar, Baez, Corichi, and Krasnov]{Ashtekar:1997yu}
A.~Ashtekar, J.~Baez, A.~Corichi, and Kirill Krasnov.
\newblock {Quantum geometry and black hole entropy}.
\newblock \emph{Phys. Rev. Lett.}, 80:\penalty0 904--907, 1998.
\newblock \doi{10.1103/PhysRevLett.80.904}.

\bibitem[Li et~al.(2018)Li, Singh, and Wang]{Li:2018opr}
Bao-Fei Li, Parampreet Singh, and Anzhong Wang.
\newblock {Towards Cosmological Dynamics from Loop Quantum Gravity}.
\newblock \emph{Phys. Rev. D}, 97\penalty0 (8):\penalty0 084029, 2018.
\newblock \doi{10.1103/PhysRevD.97.084029}.

\bibitem[Assanioussi et~al.(2018)Assanioussi, Dapor, Liegener, and Paw{\l}owski]{Assanioussi:2018hee}
Mehdi Assanioussi, Andrea Dapor, Klaus Liegener, and Tomasz Paw{\l}owski.
\newblock {Emergent de Sitter Epoch of the Quantum Cosmos from Loop Quantum Cosmology}.
\newblock \emph{Phys. Rev. Lett.}, 121\penalty0 (8):\penalty0 081303, 2018.
\newblock \doi{10.1103/PhysRevLett.121.081303}.

\bibitem[Assanioussi et~al.(2019)Assanioussi, Dapor, Liegener, and Paw{\l}owski]{Assanioussi:2019iye}
Mehdi Assanioussi, Andrea Dapor, Klaus Liegener, and Tomasz Paw{\l}owski.
\newblock {Emergent de Sitter epoch of the Loop Quantum Cosmos: a detailed analysis}.
\newblock \emph{Phys. Rev. D}, 100\penalty0 (8):\penalty0 084003, 2019.
\newblock \doi{10.1103/PhysRevD.100.084003}.

\bibitem[Giesel et~al.()Giesel, Liu, and Rullit]{tba}
Kristina Giesel, Hongguang Liu, and Eric Rullit.
\newblock {Investigation of the gravitational collapse of the LQG-inspired effective asymmetric bounce model}.
\newblock to appear.

\bibitem[Hassan and Husain(2017)]{Hassan:2017cje}
Syed~Moeez Hassan and Viqar Husain.
\newblock {Semiclassical cosmology with polymer matter}.
\newblock \emph{Class. Quant. Grav.}, 34\penalty0 (8):\penalty0 084003, 2017.
\newblock \doi{10.1088/1361-6382/aa6455}.

\end{thebibliography}

\end{document}